\documentclass[letters,twocolumn,usenatbib]{mnras}

\usepackage[T1]{fontenc}
\usepackage{ae,aecompl}

\usepackage{graphicx}	
\usepackage{amsmath}	
\usepackage{amssymb}	

\newcommand{\dd}{\; \mathrm{d}}
\newcommand{\ii}{\mathrm{i}}
\newcommand{\e}{\mathrm{e}}
\newcommand{\TF}{\mathcal{F}}
\newcommand{\M}{\mathcal{M}}
\newcommand{\Id}{\mathrm{Id}}

\title[Coronagraphic imaging through turbulence]{An analytic expression for coronagraphic imaging through turbulence. Application to on-sky coronagraphic phase diversity.}

\author[O. Herscovici-Schiller, L. M. Mugnier and J.-F. Sauvage]{
Olivier Herscovici-Schiller$^1$\thanks{E-mail: olivier.herscovici@onera.fr (OHS)},
Laurent M. Mugnier$^1$
and Jean-Fran\c{c}ois Sauvage$^{1,2}$
\\
$^{1}$ONERA -- The French Aerospace Lab, 92322 Ch\^atillon, France\\
$^{2}$Aix-Marseille Universit\'e, CNRS, Laboratoire d'Astrophysique de Marseille UMR 7326, 13388 Marseille, France
}

\date{Accepted XXX. Received YYY; in original form ZZZ}

\pubyear{2017}

\begin{document}
\label{firstpage}
\pagerange{\pageref{firstpage}--\pageref{lastpage}}
\maketitle

\begin{abstract}
The ultimate performance of coronagraphic high contrast exoplanet imaging
systems such as SPHERE or GPI is limited by quasi-static aberrations. These
aberrations produce speckles that can be mistaken for planets in the image. 
In order to design instruments, correct quasi-static aberrations or analyse
data, the expression of the point spread function of a coronagraphic instrument
in the presence of residual turbulence is most useful. Here we derive an
analytic expression for this point spread function that is an extension to coronagraphic imaging of Roddier's expression for imaging through turbulence.
We give a physical interpretation of its
structure, we validate it by numerical simulations and we show that it is
computationally efficient. Finally, we incorporate this imaging model into a coronagraphic phase diversity method (COFFEE) and validate by simulations that it allows wave-front reconstruction in the presence of residual turbulence. The preliminary results, which give a sub-nanometric precision in
the case of a SPHERE-like system, strongly suggest that quasi-static
aberrations could be calibrated during observations by this method. 
\end{abstract}





\section{Introduction}

Direct detection of exoplanets is limited by the presence of speckles on scientific images. These speckles are induced by quasi-static aberrations in the optical system that are not corrected by the adaptive optics loop. On SPHERE on the VLT, the quasi-static aberrations can be calibrated on an internal source thanks to COFFEE (see \cite{Paul-a-14}), an extension of phase diversity (see \cite{Gonsalves-82} for the principle of phase diversity, and \cite{Mugnier-l-06a} for a review) to coronagraphic imaging. However, the quasi-static aberrations evolve during the observations, so it would be useful to calibrate them during the observations in order to reach the ultimate performance of the instrument. Since COFFEE uses a model of the instrument, an analytic expression of the long exposure point spread function of a coronagraphic instrument in the presence of the residual turbulence of an adaptive optics system is needed.

To the best of our knowledge, no such general analytic expression has been previously published, even though the shape and properties of the coronagraphic point spread function have been explored, e.g. in \cite{Perrin03,Soummer-apj-07}, and an analytic approach has previously been developed in~\cite{Sauvage10}. However, \cite{Sauvage10} was based on the hypothesis of a perfect coronagraph, which does not exist, and yields unacceptable errors when used for coronagraphic phase diversity, as shown by \cite{Paul13aa}.

The purpose of this letter is firstly to present a general analytic expression for coronagraphic imaging through turbulence -- which is an extension to the coronagraphic case of the well-known expression by \cite{Roddier81} for imaging through turbulence -- and secondly to show that it can be integrated into COFFEE to measure quasi-static aberrations with nanometric precision during observations.

Moreover, the expression that we derive in this letter could be of great use for at least two other applications than wave-front sensing. One is the design, simulation and optimisation of coronagraphic systems. The other is the processing of high contrast images based on an imaging model, as proposed by \cite{Ygouf-a-13}, which is essential for the detection and characterisation of exoplanets.

\section{Classical imaging through turbulence and instantaneous coronagraphic point spread function}
\subsection{Classical imaging through turbulence}
Let us start with a reminder of the non-coronagraphic case of imaging through turbulence. The Fourier transform of Equation~(4.15) of the classical text by \cite{Roddier81} shows that the long exposure point spread function $h_{le}$ of a telescope through a (possibly residual) turbulent atmosphere is 
\begin{equation}
\label{eq:Roddier}
 h_{le} = h_a \star h_s,
\end{equation}
where $h_s$ is the point spread function of the optical system (which accounts for quasi-static aberrations), $\star$ is the convolution operator, and $h_a$ is the so-called atmospheric point spread function, which is the inverse Fourier transform of the atmospheric transfer function $\tilde{h}_a$:
\begin{equation}
\label{eq:h_a}
\tilde{h}_a =  \exp\left(-\frac{1}{2} D_\phi\right),
\end{equation}
$D_\phi$ being the phase structure function of the (possibly residual) atmospheric turbulence.

This expression is appealing because it untangles the influence of the atmospheric turbulence from the influence of the quasi-static aberrations on the global point spread function.

We show below that an analogous formula can be derived for coronagraphic systems.

\subsection{Instantaneous point spread function of a telescope equipped with a coronagraph}
\begin{figure}
\begin{center}
\includegraphics[width=0.45\textwidth]{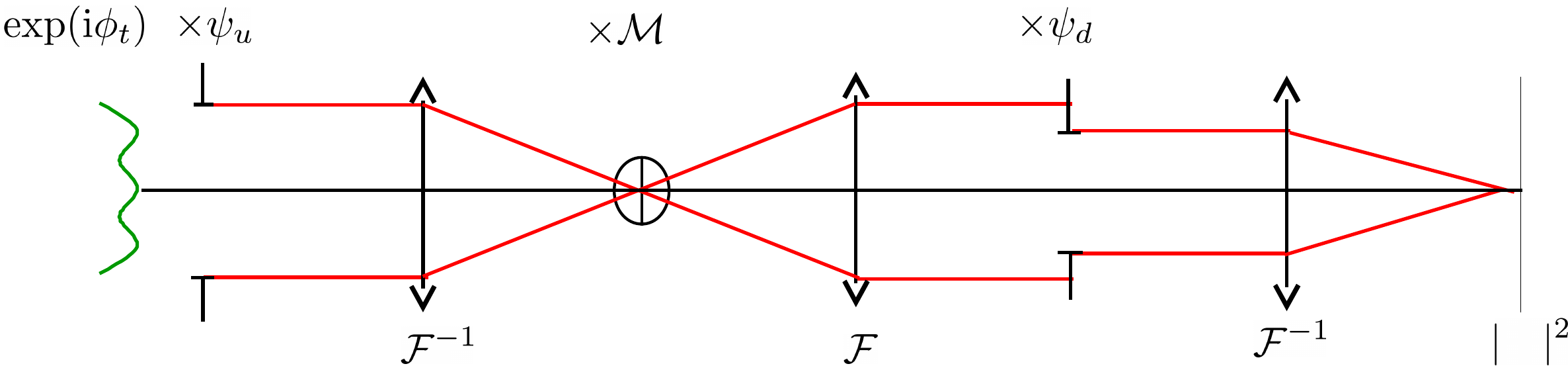} 
\caption{Sketch of a telescope equipped with a coronagraph}
\label{telescope}
\end{center}
\end{figure}

	Let us consider an imaging system on a telescope equipped with a coronagraph as described in
    Figure~\ref{telescope}. We will denote by $P_u$ the amplitude transmission
    in a pupil plane before the coronagraph, or ``upstream pupil'', $\phi_u$
    the phase aberrations in this plane, which will be called ``upstream
    aberrations'' hereafter, and $\psi_u = P_u \e^{\ii \phi_u}$ the complex
    amplitude in this plane. Similarly, we will denote by $P_d$ the downstream
    transmission in a pupil plane after the coronagraph, or ``downstream
    pupil'' (also called the Lyot stop plane), $\phi_d$ the phase aberrations in this plane, which will be
    called ``downstream aberrations'' hereafter, and $\psi_d = P_d \e^{\ii
      \phi_d}$ the multiplicative term -- due to the downstream aberrations -- that affects the complex amplitude of the wave that has propagated from the entrance aperture to this plane.  
      For the sake of simplicity, we want to keep the same orientation of the images from one focal plane to the following, and the same pupil image orientation from one pupil plane to another. 
      For this reason, we rotate the axis of the downstream plane by 180° with respect to the upstream plane, and similarly we rotate the orientation of the focal mask plane by 180° with respect to the final focal plane. 
This axis change is taken into account mathematically by choosing an inverse Fourier transform ($\TF^{-1}$) instead of a forward one ($\TF$) to model the propagation of light from a pupil to a focal plane. 
  This convention allows for the image orientation to always remain the same. Of course, our main result (Equation (\ref{eq:physique})) is independent of this orientation convention. 

Let us denote by $h_c$ the coronagraphic point spread function of the instrument without turbulence. It writes
\begin{equation}
		h_c\left(\alpha;\psi_u ,\psi_d\right) =   \left|\TF^{-1}\left\lbrace \psi_d \times \TF \left[ \M \times \TF^{-1}\left( \psi_u  \right) \right] \right\rbrace (\alpha) \right|^2, 
\end{equation}
where $\alpha$ is the vector of angular coordinates of a position on the sky, and $\M$ the coronagraphic focal plane mask. This equation simply expresses the propagation from successive planes to the next ones and the quadratic detection, illustrated in Figure~\ref{telescope}.
	For notation simplicity we take $\lambda = 1$ in the equations, that is to say we use the same coordinate system for the pupil phases and for the transfer functions.

	In the presence of turbulence, the instantaneous point spread function of this instrument writes
	\begin{equation}
		h_{sec}(\alpha,t ;\psi_u,\psi_d) = h_c(\alpha; \psi_u \times \e^{\ii\phi_t(t)},\psi_d).
	\end{equation}
	Here $h_{sec}(\alpha,t,\psi_u,\psi_d)$ ($sec$ stands for ``short exposure coronagraphic'') is the instantaneous point spread function taken at coordinate $\alpha$ in the detector focal plane at time $t$ and $\phi_t(t)$ is the instantaneous residual turbulence phase at time $t$ if the telescope is ground-based and equipped with adaptive optics.
	
\section{Long exposure coronagraphic point spread function}
In current high contrast systems such as SPHERE or GPI, the exposure time is large with respect to the typical evolution time of corrected turbulence, so that the signal that is actually recorded by the detector is the long exposure point spread function, which is an average over time of the short exposure point spread function:
	\begin{equation}
		h_{lec}(\alpha ; \psi_u,\psi_d) = \left\langle h_{sec}(\alpha,t;\psi_u,\psi_d) \right\rangle_t .
	\end{equation}		
	Here $h_{lec}$ is the long exposure coronagraphic point spread function, and $\left\langle  \right\rangle_t$ is the averaging over time. Let us denote by $f^*$ the complex conjugate of any function $f$, and by $\otimes$ the bi-dimensional correlation product:
    $$ f\otimes g (r) = \iint f^*(s)g(r+s) \dd s.  $$
    
	In developed form, the long exposure coronagraphic point spread function writes, thanks to Wiener-Khintchine's theorem:
\begin{equation}
    \begin{split}
		h_{lec}(\alpha;\psi&_u,\psi_d) = \\
       & \left\langle \TF^{-1}\left(\left\lbrace \psi_d \times \TF \left[ \M \times \TF^{-1}\left( \psi_u \e^{\ii\phi_t(t)} \right) \right]\right\rbrace \right. \right. \\
       & \otimes \left. \left. \left\lbrace \psi_d \times \TF \left[ \M \times \TF^{-1}\left( \psi_u \e^{\ii\phi_t(t)} \right) \right]  \right\rbrace \right) (\alpha) \right\rangle_t.
	\end{split}
    \end{equation}
	
	To somewhat simplify calculations, we will consider the optical transfer function instead of the point spread function. Let $\tilde{h}_{lec}~=~\TF(h_{lec})$ be the Fourier transform of the point spread function.
	The integral over space commutes with the averaging over time, so 
	\begin{equation*}
    \begin{split}
		\tilde{h}_{lec}(r;\psi&_u,\psi_d)  = \\
        & \left\langle \left\lbrace \psi_d \times \TF \left[ \M \times \TF^{-1}\left( \psi_u \e^{\ii\phi_t(t)} \right) \right]\right\rbrace \right. \\
        & \left.  \otimes \left\lbrace \psi_d \times \TF \left[ \M \times \TF^{-1}\left( \psi_u \e^{\ii\phi_t(t)} \right) \right]  \right\rbrace (r) \right\rangle_t.
    \end{split}
	\end{equation*}
	In order to express the average over time, we must fully develop this expression. We introduce the pupil-plane variables $r_1$, $r_2$ and $r_3$, the focal-plane variables $\alpha_1$ and $\alpha_2$. After some manipulations, the fully developed expression reads
	\begin{equation*}	
	\begin{split}
		\tilde{h}_{lec}(r;\psi_u,\psi_d) = &\iint \psi_d^*(r_1) \psi_d(r+r_1) \iint \iint \e^{\ii 2 \pi r_1 \cdot \alpha_1} \\
        & \times \e^{-\ii 2 \pi (r_1+r) \cdot \alpha_2} \M^*(\alpha_1)\M(\alpha_2) \\
        & \times \iint \iint \e^{-\ii 2 \pi r_2 \cdot \alpha_1}\e^{\ii 2 \pi r_3 \cdot \alpha_2} \psi_u^*(r_2) \psi_u(r_3)  \\
        & \times \left\langle \e^{\ii\left[ \phi_t(r_3,t) - \phi_t(r_2,t) \right]} \right\rangle_t \dd r_3 \dd r_2 \dd \alpha_2 \dd \alpha_1 \dd r_1 .
	\end{split}
	\end{equation*}
	Following \cite{Roddier81}, we assume that turbulence is an ergodic stationary process (which is a very reasonable assumption in the case of residual turbulence after an extreme adaptive optics system), so we write 
	\begin{equation}
	\left\langle \e^{\ii\left[ \phi_t(r_3,t) - \phi_t(r_2,t) \right]} \right\rangle_t = \e^{-\frac{1}{2}D_\phi(r_3-r_2)},
\end{equation}
where $D_\phi$ is the turbulent phase structure function, defined by 

\begin{equation}
D_\phi(r) = \left\langle \e^{\ii\left[ \phi_t(r',t) - \phi_t(r+r',t) \right]} \right\rangle_t.
\label{Dphi}
\end{equation}

In order to be able to separate variables $r_2$ and $r_3$ in Equation~(\ref{Dphi}), we take the inverse 
 Fourier transform of $ \e^{-\frac{1}{2}D_\phi}$. We denote $\alpha'$ the conjugate variable of $r_3-r_2$, and  $h_a$ is the atmospheric point spread function defined by Equation~(\ref{eq:h_a}). We will come back to the meaning of $h_a$ in Section~\ref{sec:physical_interpretation}. We obtain:
\begin{equation*}	
	\begin{split}
		\tilde{h}_{lec}(r;\psi&_u,\psi_d,D_\phi) = \\
        & \iint h_a(\alpha';D_\phi) \iint \psi_d^*(r_1) \iint \e^{\ii 2 \pi r_1 \cdot \alpha_1} \M^*(\alpha_1) \\ 
        &\times \iint  \e^{-\ii 2 \pi r_2 \cdot \alpha_1} \psi_u^*(r_2) \e^{-\ii 2 \pi r_2 \cdot \alpha'} \dd r_2 \dd \alpha_1 \\ 
		& \times  \psi_d(r+r_1)  \iint  \e^{-\ii 2 \pi (r_1+r) \cdot \alpha_2} \M(\alpha_2) \\
        & \times \iint \e^{\ii 2 \pi r_3 \cdot \alpha_2}  \psi_u(r_3)  \e^{\ii 2 \pi r_3 \cdot \alpha'}  \dd r_3  \dd \alpha_2  \dd r_1 \dd \alpha' .
	\end{split}
	\end{equation*}
In a more compact form, using $\Id$ to denote the identity function, this equation can be written:
\begin{equation}
\begin{split}
		\tilde{h}_{lec}(r;\psi&_u,\psi_d,D_\phi) =  \\
		& \iint h_a(\alpha';D_\phi)  \left\lbrace \psi_d \times \TF \left[ \M \times \TF^{-1}\left( \psi_u \e^{\ii 2 \pi \alpha' \cdot \Id)} \right) \right]\right\rbrace \\
        & \otimes \left\lbrace \psi_d \times \TF \left[ \M \times \TF^{-1}\left( \psi_u \e^{\ii 2 \pi \alpha' \cdot \Id)}\right) \right]  \right\rbrace (r)  \dd \alpha'.
		\end{split}
	\end{equation}

To obtain the point spread function back from this optical transfer function, we just have to take the inverse Fourier transform, then apply Wiener-Khintchine's theorem again. Finally, the long exposure coronagraphic point spread function reads 
\begin{equation}
\label{eq:psf}
\begin{split}
		h&_{lec}(\alpha;\psi_u,\psi_d,D_\phi) =  \\ 
        &\iint h_a(\alpha';D_\phi)  \left|\TF^{-1}\! \left\lbrace \psi_d . \TF \left[ \M . \TF^{-1}\! \left( \psi_u \e^{\ii 2 \pi \alpha' \cdot \Id} \right) \right] \right\rbrace \! (\alpha) \right|^2   \dd \alpha'. 
        \end{split}
	\end{equation}

This expression gives the long exposure coronagraphic point spread function as a function of three deterministic parameters, namely: upstream aberrations, downstream aberrations, and residual turbulence-induced phase structure function.

\section{Physical interpretation}
\label{sec:physical_interpretation}
In order to interpret Equation (\ref{eq:psf}) physically, we can re-write it in the more compact following form:
\begin{equation}
	\label{eq:physique}
    \boxed{
    \begin{split}
    		h_{lec}&(\alpha;\psi_u,\psi_d,D_\phi) =  \\
            & \iint h_a(\alpha';D_\phi) \; h_c\!\left(\alpha;\psi_u \e^{\ii 2 \pi \alpha' \cdot \Id },\psi_d\right)  \dd \alpha' ,
    \end{split}
}
\end{equation}
    
where $h_c\!\left(\alpha;\psi_u \e^{\ii 2 \pi \alpha' \cdot \Id },\psi_d\right)$ is the coronagraphic point spread function of the instrument in the absence of turbulence, but with a tilt $\alpha'$ \emph{added} to the upstream aberrations, that is to say with the light coming from the star being tilted by an angle $\alpha'$.

A physical interpretation of the atmospheric point spread function defined in Eq. (\ref{eq:h_a}) can shed some light on the expression of Eq.~(\ref{eq:physique}). 
Recall that
\begin{equation*}
h_a \triangleq \TF^{-1}\left[ \exp\left(-\frac{1}{2} D_\phi\right) \right] = \TF^{-1}\left[\left\langle \e^{\ii\left[ \phi_t(r+r',t) - \phi_t(r',t) \right]} \right\rangle_t \right] .
\end{equation*}

If we note $\psi_t(r,t) = \exp\left( \ii \phi_t(r,t) \right)$ the contribution to the electric field at position $r$ and time $t$ by the atmospheric turbulence, we can re-write $h_a$ as :
\begin{equation}
h_a = \TF^{-1}\left[\left\langle \psi_t^*(r',t)\psi_t(r+r',t)  \right\rangle_t \right]
\end{equation}
Assuming stationariness and ergodicity, we recognize the inverse Fourier transform of the auto-correlation of $\psi_t$. Hence we can identify $h_a$ as the energy spectral density of the turbulence-induced complex field. In other words, $h_a(\alpha')$ gives the fraction of the energy of light that is diffracted in each direction $\alpha'$. 

Finally, Equation~(\ref{eq:physique}) can be interpreted as an incoherent plane wave decomposition: the long exposure coronagraphic point spread function $h_{lec}$ is a weighted sum of coronagraphic point spread functions $h_c$, without any turbulence, but with an upstream tilt $\alpha'$. The weight on any of these tilted point spread functions is the fraction of the light energy $h_a(\alpha')$ that the atmosphere scatters in the direction of the corresponding tilt $\alpha'$.

Moreover, the expression of Equation (\ref{eq:physique}) separates the turbulent part and the coronagraphic part of the problem: $h_a$ codes for the characteristics of the turbulent atmosphere, while $h_c$ codes for the characteristics of the instrument. This separation of the point spread function into an atmospheric point spread function and a point spread function of the instrument is strikingly similar to the one found in the Roddier expression for non-coronagraphic imaging.

\section{Special cases}
\subsection{Non-turbulent point spread function}
As a first particular case, let us assume that there is no turbulence, so that $h_a$ is reduced to a Dirac distribution.
Indeed, in the case where there is no turbulence, $\phi_t$ is a constant, so $D_\phi = \left\langle \left[ \phi_t(r') - \phi_t(r+r')\right]^2 \right\rangle_{r'} = 0$. Then, according to Equation (\ref{eq:h_a}), $h_a = \delta.$
Hence, by use of Equation (\ref{eq:physique}),
\begin{equation}
		h_{lec}(\alpha;\psi_u,\psi_d)  =  h_c(\alpha;\psi_u,\psi_d),
	\end{equation}
which is precisely the classical expression for a coronagraphic point spread function in the absence of turbulence.

\subsection{Non-coronagraphic optical transfer function}
As a second particular case, if we now consider a non-coronagraphic instrument, $\M = 1$, so Equation (\ref{eq:psf}) reads 
\begin{equation}
\begin{split}
		h_{lec}(\alpha; \psi&_u,\psi_d,D_\phi) \\
        & = \iint h_a(\alpha';D_\phi)  \left|\TF^{-1}\left\lbrace \psi_d   \psi_u \e^{\ii 2 \pi \alpha' \cdot \Id}  \right\rbrace (\alpha) \right|^2   \dd \alpha' \\
        & = h_a(D_\phi) \star \left|\TF^{-1}\left\lbrace \psi_d   \psi_u \right\rbrace\right|^2 (\alpha), 
        \end{split}
	\end{equation}
which is identical to Equation~(\ref{eq:Roddier}), with $h_s = \left| \TF^{-1}\left\lbrace \psi_d   \psi_u \right\rbrace \right|^2$, that is to say that the aberration is the sum of the upstream and downstream aberrations.
The long exposure optical transfer function is thus
\begin{equation}
		\tilde{h}_{lec}(r;\psi_u,\psi_d)  = \tilde{h}_a(r) \times \left(\psi_d \psi_u \otimes \psi_d \psi_u\right) (r),
	\end{equation}
which is the classical formula by \cite{Roddier81} of the long exposure transfer function for imaging through turbulence; indeed the autocorrelation of $\psi_d \psi_u$ is the optical transfer function of the instrument due to diffraction and quasi-static aberrations.

\subsection{Approximation in the case of weak turbulence}
Additional insight on Equation (\ref{eq:physique}) can be gained by assuming that the residual turbulence phase after the adaptive optics is a small perturbation. It is straightforward to see that
\begin{equation*}
		\frac{1}{2}D_\phi(r)  = \sigma_\phi^2 -  A_\phi(r),
	\end{equation*}
 where $A_\phi(r) = \left\langle  \phi_t(r')  \phi_t(r+r') \right\rangle_{r'}$ is the autocorrelation of the phase, hence
\begin{equation*}
		h_a = \e^{-\sigma_\phi ^2} \TF^{-1}\left[ \e^{A_\phi} \right]. 
	\end{equation*}
Then, if the turbulence is small, $A_\phi$ is small compared to $1$, and we can perform a first order Taylor expansion:
\begin{equation*}
\begin{split}
		h_a(\alpha') & \simeq \e^{-\sigma_\phi ^2} \TF^{-1}\left[ 1+ A_\phi \right](\alpha') \\
		h_a(\alpha') & \simeq \e^{-\sigma_\phi ^2} \delta(\alpha') + \e^{-\sigma_\phi ^2} S_\phi(\alpha')
\end{split}
	\end{equation*}
where $S_\phi$ is the power spectrum density of the turbulent phase $\phi_t$, defined as the (inverse) Fourier transform of $A_\phi$. Thus we can express $h_{lec}$ :
\begin{equation}
\begin{split}
		h&_{lec}(\alpha;\psi_u,\psi_d) \simeq \\
        &\e^{-\sigma_\phi ^2} \times \left[ h_c(\alpha;\psi_u,\psi_d) + \iint S_\phi(\alpha') h_c\left(\alpha;\psi_u e^{\ii 2 \pi \alpha '},\psi_d\right)  \dd \alpha' \right]
        \end{split}
	\end{equation}
This means that $h_{lec}$ is approximately the non-turbulent coronagraphic point spread function, with a corrective additive term that takes into account the power spectrum density of the turbulent phase, all this dampened by the coherent energy $\e^{-\sigma_\phi^2}$.

\section{Numerical validation and efficiency considerations}
\subsection{Numerical validation}
\begin{figure*}
\begin{center}
\includegraphics[width=0.8\textwidth]{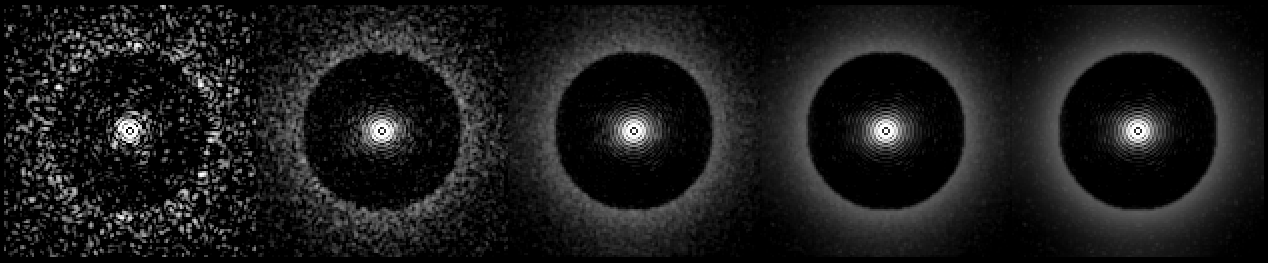} 
\caption{Point spread functions. From left to right: short exposure, averages of 10, 100 and 1000 exposures, and analytic long exposure calculated with equation~\ref{eq:physique}.}
\label{conv_psfs}
\end{center}
\end{figure*}

To validate our point spread function model, we test it against an average of short exposure point spread functions, each one of them with a different outcome of the residual turbulence-induced phase. Each short exposure point spread function is computed using a matrix Fourier transform as proposed by \cite{Soummer-opex-07} in order to perform an accurate computation of the field on the mask focal plane.
A few averages of point spread functions are displayed on Figure~\ref{conv_psfs}. They are simulated with a Lyot coronagraph of diameter $3\lambda/D$, the residual turbulence of a SPHERE-like adaptive optics and upstream and downstream white noise phase aberrations of variance $0.1\; \textrm{rad}^2$.

Let us define the convergence error as follows. We denote by $h_{lec}$ the long exposure point spread function using the analytic formula of Equation (\ref{eq:physique}). We denote $h_{sseN}$ the average of $N$ short exposures. We define the error as $\mathrm{err}_{N} = \mathrm{max}\left(\left| h_{sseN} - h_{lec} \right| / h_{lec}\right),$ where the division is taken pixel by pixel. We have performed a simulation of the evolution of error versus $N$, the number of short exposures. The error tends to zero as $N^{-1/2}$, which is consistent with the central limit theorem. The error goes below $10^{-3}$ just after $N = 10^7$ short exposures.
 
\subsection{Computing cost}
The evolution of the convergence error gives us an easy criterion to quantify the comparative computing costs of the long exposure coronagraphic point spread function and the empirical average of short exposures.
Let us take the computing cost of a short exposure point spread function as the unit computing cost. Then, if we want an error of less than $10^{-3} $ on the point spread function, approximately $10^7$ short exposure point spread functions must be averaged, for a cost of $10^7$. The analytic formula for the long exposure point spread function has a total cost of the number of points on which the phase structure function is known. This implies that, for square images of $512 \times 512$ pixels, our exact formula is about 38 times less costly to evaluate than an average of short exposures.

In addition, it should be noted that, since the long exposure is an integral, and thus, in practice, a sum, it is very easy to compute it in parallel on several processors. We made all calculations in parallel on 16 cores.

Finally, it should also be noted that a useful approximation can accelerate the computing of the long exposure point spread function by a great deal. Indeed, when the tilt $\alpha'$ is sufficiently greater than the radius of the focal-plane mask of the coronagraph, the point spread function can be well approximated by a shifted non-coronagraphic point spread function that can be computed once and for all, so the sum in Equation (\ref{eq:physique}) for the computation of the long exposure point spread function must actually only be computed on a square of typical width $8\lambda/D$. For a Nyquist-rate sampled image, that is only $16\times16$ short exposure point spread functions, with a maximum relative error of less than $10^{-3}$ on each pixel for SPHERE-like parameters.

\section{Application: estimation of quasi-static aberrations  with COFFEE in the presence of residual turbulence}
	\subsection{Motivation}
	In current systems such as SPHERE, quasi-static aberrations are only corrected during day-time. For an upgrade, or for future high contrast instruments on extremely large telescopes, it would be useful to correct them during the scientific acquisition.
	To achieve this, we combine COFFEE (see \cite{Paul-a-13b}, \cite{Paul-a-14}) with the long exposure phase diversity proposed by \cite{Mugnier-a-08}.

\subsection{Simulation results}
\begin{figure}
\centering
\begin{minipage}{.75\linewidth}
  \centering
  \includegraphics[width=1\linewidth]{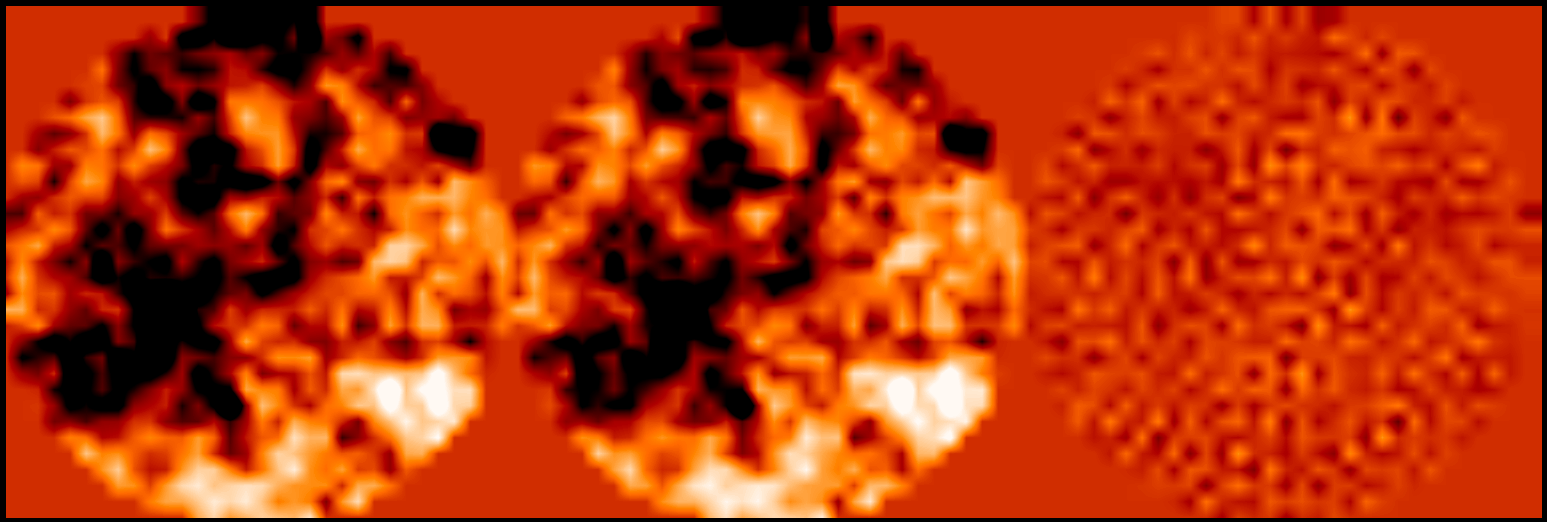}
  \label{fig:test1}
\end{minipage}%
\begin{minipage}{.25\linewidth}
  \centering
  \includegraphics[height=1\linewidth]{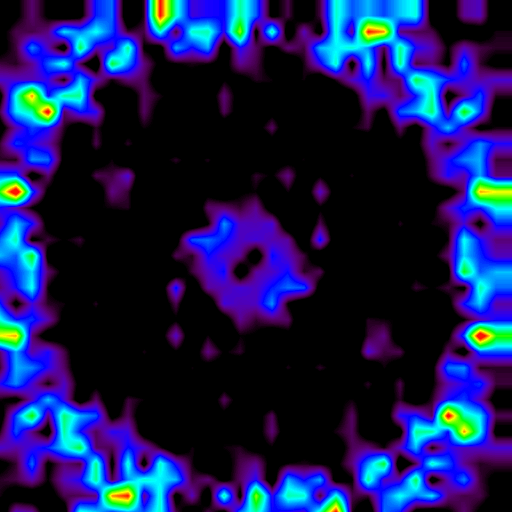}
  \label{fig:test2}
\end{minipage}
\caption{Left: upstream aberrations that we aim to reconstruct. Center left: reconstructed aberration. Center right: difference, magnified a hundred times. Right: modulus of the Fourier transform of the phase estimation error}
\label{sortie_coffee}
\end{figure}

We have incorporated the long exposure coronagraphic imaging model developed above into the coronagraphic phase diversity method COFFEE. In the following, we perform a preliminary validation of the quasi-static aberration estimation with COFFEE in the presence of residual turbulence by means of simulation. The parameters of the simulations are the following:
the phase map size is $64 \times 64$ pixels, the incoming flux is $10^9$ photoelectrons, the standard deviation of the readout noise is 1 electron, the (supposedly monochromatic) wavelength is 1589 nm. The phase structure function is typical of the residual turbulence of the extreme adaptive optics system of the SPHERE instrument of the VLT, the coronagraph is a Roddier \& Roddier phase mask, and there are no downstream aberrations.
The upstream aberration is 50 nm RMS.
Figure \ref{sortie_coffee} shows the true upstream aberrations, the estimated upstream aberrations, and the difference between the upstream aberrations and the estimated ones, magnified a hundred times. The error in this case is much less than 1 nm RMS, which is a very encouraging result. 
Additionally, the usefulness of this reconstruction is even better than what the estimated error suggests. Indeed, if we examine the Fourier transform of the difference on Figure~\ref{sortie_coffee}, we notice that the error is mainly located on the spatial frequencies that are on the border of the array, that is to say, those that are not corrected by the deformable mirror. Quantitatively, the error in the corrected zone accounts for less than 10 \% of the total estimation error variance.

\section{Conclusion and perspectives}
To summarize this work, we have derived an analytic expression to model long exposure image formation for a coronagraphic telescope through the residual turbulence of an extreme adaptive optics system. This model is general and valid for any coronagraphic mask. It only depends on three deterministic quantities: quasi-static aberrations before and after the coronagraph, and the residual phase structure function. Its expression shows, quite nicely, that the long exposure coronagraphic point spread function is the weighted sum of coronagraphic point spread functions without turbulence with a weight distribution given only by the turbulence-induced residual phase structure function.  It is computationally effective, allowing a time gain of at least an order of magnitude on standard sized images.
Lastly, we have obtained numerical results that suggest that this long exposure model can be applied to the on-sky measurement and correction of aberrations in the context of ground-based direct exoplanet detection.

These encouraging results suggest that we carry on working in two directions. On the one hand, we will explore the robustness of the COFFEE estimations to an error on the phase structure function of the residual atmospheric turbulence. On the other hand, we will demonstrate experimentally that COFFEE can estimate aberrations through turbulence, first on a laboratory bench and then on sky.

 \section*{Acknowledgements}
 The PhD work of O. Herscovici-Schiller is co-funded by CNES and ONERA. We thank J.-M. Le Duigou for his support.
This work received funding from the E.U. under FP7 Grant Agreement No. 312430
OPTICON, from the CNRS (D\'efi Imag'In), and from ONERA in the framework of
the VASCO research project.  We thank Dr Faustine Cantalloube for useful
discussions. We thank the reviewer for his careful review and constructive comments.




\bibliographystyle{mnras}
\bibliography{biblioOHS} 
\bsp	
\label{lastpage}
\end{document}